\documentclass[conference]{IEEEtran}
\usepackage{graphicx}
\usepackage{adjustbox}

\ifCLASSINFOpdf
\else
\fi
\begin{document}
%
\title{On the  Differential Cryptanalysis of SEPAR Cipher}

\author{\IEEEauthorblockN{Arsalan Vahi}
\IEEEauthorblockA{Department of computer engineering\\
Science and research branch\\
Islamic Azad university, Tabriz, Iran\\
Email: arsalan.vahi2009@gmail.com}
\and
\IEEEauthorblockN{Mirkamal Mirnia}
\IEEEauthorblockA{Department of Applied Mathematics\\
and Computer Science\\
University of Tabriz, Tabriz, Iran\\
Email: mirnia-kam@tabrizu.ac.ir}
}


%


\maketitle

\begin{abstract}
 SEPAR is a lightweight cryptographic algorithm, designed to implement on resource-constrained devices especially those employed in IoT environments. Meanwhile, the mixed structure design of cipher leads to speed improvement while guaranteeing its resistance against common cryptographic attacks, especially differential and linear attacks. In order to confirm the resistance of the cipher against differential attack, an extensive investigation was presented in our previous work. In his study, we conduct new research continuing the previously presented research. We prove that there are enough active S-boxes so as to resist cipher against differential cryptanalysis. Moreover, this can provide a tight bound of resisting cipher against this attack. 
 

\end{abstract}


%
\IEEEpeerreviewmaketitle

\section{Introduction}
SEPAR is a new lightweight cryptographic algorithm targeted for low-constrained devices. The hybrid design structure of this cipher is a smart combination of a Finite State Machine (FSM) and Block cipher primitive, which is motivated by the design scheme of Axel York Poschmann \cite{vahi2020separ}. This idea leads to taking advantage of stream cipher speed, meanwhile utilizing appropriate resistance of block ciphers against common attacks. High security and resistance against cryptographic attacks are necessary for applying a cipher in real-world applications; therefore, the resistance of SEPAR was investigated in a vast range of attacks including birthday attack, algebraic attacks, related-key attacks, collision attacks, etc. Also, Linear and Differential cryptanalysis \cite{heys2002tutorial} of SEPAR was investigated extensively in the proposal paper\cite{vahi2020separ}. 

 Differential cryptanalysis analyzes pairs of plaintexts and seeks to exploit how the difference between these plaintexts propagates through a block cipher. Typically, reduced round variants of block ciphers are analyzed, since analyzing an entire modern block cipher is typically computationally infeasible \cite{jithendra2020new}. 

In the case of differential cryptanalysis regarding the Present cipher, the presented theory and practical confirmation by \cite{poschmann2009lightweight} prove that any differential characteristic over 25 rounds must have at least 50 active S-Boxes. The maximum differential probability of a present S-Box is $2^{-2}$ and so the probability of a single 25-rounded differential characteristic is bounded by $2^{-100}$. The bound on the number of active S-boxes was proven tight by provided practical confirmations \cite{poschmann2009lightweight}. In the same way, the BORON cipher presented a minimum of eight active S-Boxes for three rounds. As a result, in 18 rounds, there will be a minimum of 48 active S-boxes; hence, the total differential probability is $2^{-96}$. Therefore, the complete number of rounds of the BORON cipher shows good resistance against differential attacks \cite{bansod2017boron}.

Recalling that differential cryptanalysis is one of the major cryptanalysis techniques, the motivation of this paper is to focus on this attack and conduct more investigation regarding cipher's resistance against this attack by counting the number of active S-boxes.  We define a low tight bound for the SEPAR's resistance against differential attack, moreover, we show that the existence of eight Enc-blocks and employing four consecutive b16 assures its resistance against this cryptanalysis technique. The structure of this article is as follows: An overview of the SEPAR cryptographic algorithm is presented in section 2, in section 3 differential complexity of SEPAR is investigated, and finally in section 4 conclusion provided.  

\section{THE SEPAR CRYPTOGRAPHIC ALGORITHM}
The design idea of SEPAR is a smart combination of a pseudorandom generator and a pseudorandom permutation in a way that leads to a 256-bit key, 16-bit block size, 128–bit IV, and 144-bit internal states cipher. As mentioned before, the design strategy of SEPAR is to cope with the trade-off between cost, security, and performance.
\subsection{Initialization and encryption process}
Prior to the encryption process, Initialization commences and during this process, eight 16-bit nonce values feed to the algorithm, then the algorithm runs for 4 consecutive rounds. Afterward, the provided data, which are stored in states (registers), are ready for putting to use in the encryption algorithm. The overall structures of initialization, encryption, and decryption are shown in Figure 1. 
As it is obvious, the structure of encryption and initialization algorithms are the same, and both consist of eight Enc-blocks. In addition, the encryption algorithm utilizes another component, named LFSR. The main 256-bit 
\begin{figure*}[h]
  \includegraphics[width = \textwidth, height = 10cm]{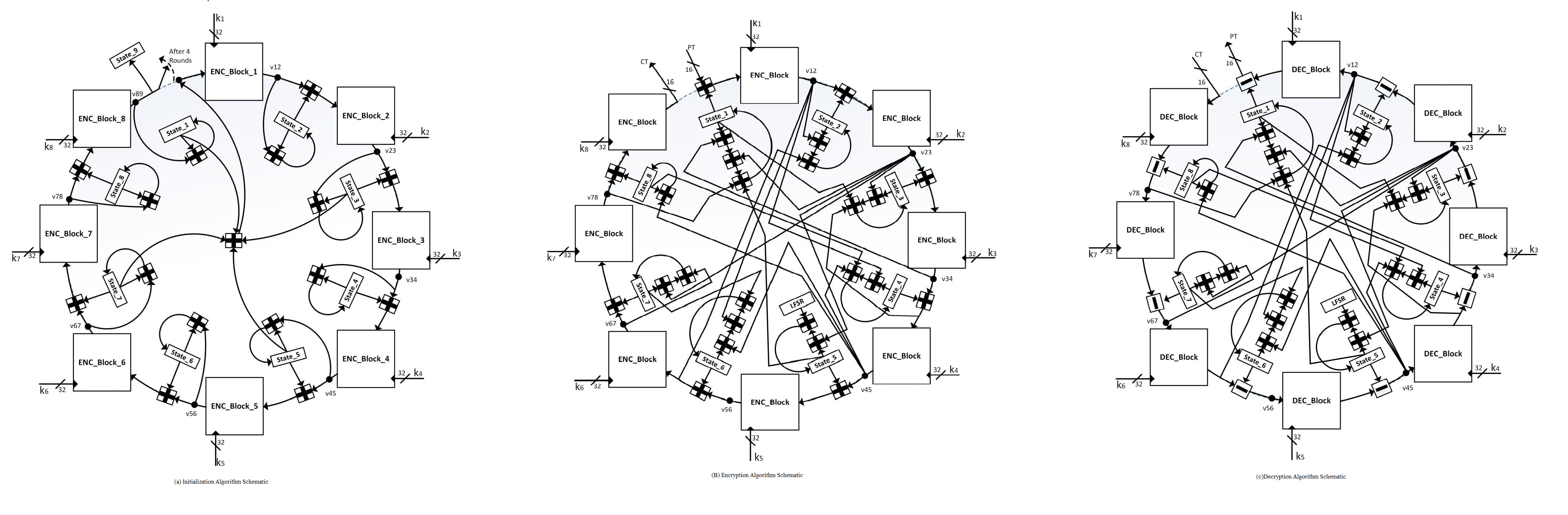}
  \caption{The SEPAR Cryptographic Algorithm and its internal structures}
\end{figure*}
  key splits into eight equal parts, named segment key that each Enc-block utilizes one. In the process of extracting sub-keys from the segment keys, which is similar to the key-generation algorithm, 6 sub-keys produce. This process not only produces non-linear sub-keys but also provides resistance against related-key and slide attacks. Moreover, by use of this method of subkey generation, no speed decrements or high overload imposed during execution. For commencing the SEPAR data encryption process, the contents of the initialization states algorithm map exactly to the contents of the encryption algorithm; also, prior to mapping state 9 contents to LFSR, the 9\textsuperscript{th}- bit sets to one. After these steps, the encryption algorithm is ready and 16-bit plaintext goes through eight Enc-blocks and respective 16-bit ciphertext produces. Afterward, the Internal states and LFSR update and make the cipher ready to accept new 16-bit data.
\subsection{Enc-block Structure}
Each Enc-block structure is an SP network with a data length of 16-bit and a 32-bit key length. The overall structure of Enc-block and b16 is shown in Fig. 2 
 Substitution layer is provided by the use of 4-bit to 4-bit S-Boxes which meticulously chosen from \cite{leander2007classification} \cite{saarinen2011cryptographic}. These S-boxes are shown in Table 1.
 \begin{table}
\caption{Selected Four S-boxes in Hexadecimal}
\begin{adjustbox}{width=\columnwidth,center}
 \begin{tabular}{|c||c|c|c|c|c|c|c|c|c|c|c|c|c|c|c|c|c|} 
 \hline
 x &0&1&2&3&4&5&6&7&8&9&A&B&C&D&E&F \\ 
 \hline\hline
 $s_{1}$(x)&1&F&B&2&0&3&5&8&6&9&c&7&D&A&E&4 \\ 
 \hline
 $s_{2}$(x) & 6 & A & F&4&E&D&9&2&1&7&C&B&0&3&5&8 \\
 \hline
 $s_{3}$(x) &C & 2 & 6 &1&0&3&5&8&7&9&B&E&A&D&F&4 \\
 \hline
 $s_{4}$(x) & D & B & 2&7&0&3&5&8&6&C&F&1&A&4&9&E \\
  \hline
\end{tabular}
\end{adjustbox}
\end{table}

 The permutation layer consists of lightweight mathematical operations such as bitwise addition and circular shift, which leads to high-speed run and easy implementations. In SEPAR implementation, instead of using four different S-boxes, we used one S-box (with regard to maintaining security). For more information regarding SEPAR, we can refer to \cite{vahi2020separ}. 
 
\section{THE COMPLEXITY OF DIFFERENTIAL ATTACK ON SEPAR}
 
 In the previous study, security proofs relating differential cryptanalysis are provided in \cite{vahi2020separ}.  As it is shown in Table 2, the resistance of cipher against differential cryptanalysis proofed with computing high probability differential characteristics for a different number of b16. In this case, we showed that after adding 3 consecutive b16, the differential value remains constant and we get an upper bound of 22. Therefore, we concluded four consecutive b16 employed in an Enc-block, provided sufficient security against this attack.
\begin{table}
\caption{Differential property of the 16-bit Block Cipher}
\begin{adjustbox}{center}
 \begin{tabular}{|c|c|} 
 \hline
 \# of b16 in Enc-block & $max_{(a\neq0.b)} D_{b16} (a.b)$ \\ 
 \hline
 0&16370\\
 \hline
 1&1016\\
 \hline
 2&84\\
 \hline
 3&22\\
 \hline
 4&22\\
 \hline
\end{tabular}
\end{adjustbox}
\end{table}

\begin{figure*}[h]
  \includegraphics[width = \textwidth, height = 5cm]{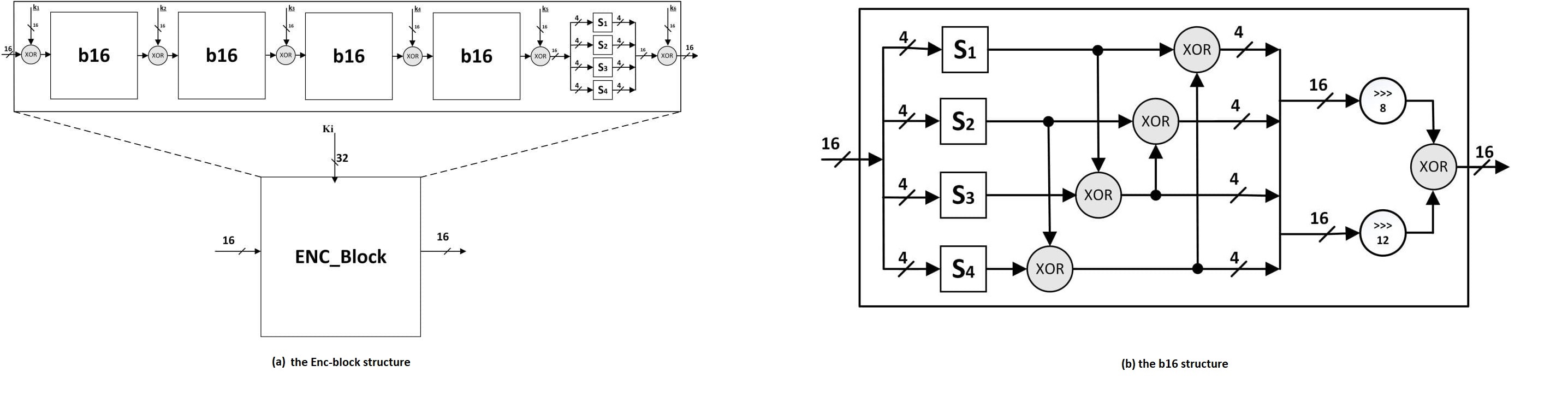}
  \caption{The overall structure of Enc-block and b16}
\end{figure*}
In this paper, in order to conduct more study relating cipher resistance against differential attack and defining a tight bound, we compute active S-box numbers in an Enc-block and offer a secure bound. Prior to this, we purpose a brief description of active S-boxes and their role in cryptanalysis.

\subsection{Active S-boxes and their roles in cryptanalysis}

Active S-boxes are those S-boxes that participate during the cryptanalysis process directly and they provide data when they feed differentials. For clarifying this definition, when  a high probability differential for example $\Delta=1101,0000,0000,0000$ is feed to a sampled encryption function, two S-boxes $S_{1}$ and $S_{2}$ participate and in other words activated during this process and we call these two ones as Active S-Box \cite{sajadieh2017new}.The purpose of computing the active S-box number is gauging the resistance of cipher against differential cryptanalysis. This gauge is a sign of a tight bound in differential cryptanalysis \cite{cui2012lower}.

 As an example, cryptanalysis of the Present cipher offers a theorem express that there is a minimum of 10 active S-boxes in any five-rounded differential characteristic of it. After providing reasonable proof, it concluded that any differential characteristic over 25 rounds of Present must have at least $10 *5 =50$ active S-boxes. The maximum differential probability of a present cipher S-box is  $2^{-2}$, therefore the probability of a single 25-round differential characteristic is bounded by $(2^{-2})^{50} = 2^{-100}$ \cite{poschmann2009lightweight}. We should mention that there is another way to compute active S-Boxes namely MILP programming wherein we compute active S-boxes numbers in SageMath environment by the use of Mixed-Integer Linear Programming \cite{sasaki2017new}.
 \subsection{Computing active S-box numbers in SEPAR}
The results of an exhaustive search to find differential characteristics with high probability for a different number of b16 are shown in Table 3. As is obvious, 7 high probability differential characteristics are obtained.
\begin{table}
\caption{ High Probably Differential characteristics for an Enc-block with 4 consecutive b16 }
\begin{adjustbox}{width=\columnwidth,center}
 \begin{tabular}{|c|c|c|} 
 \hline
 \#  & Input Differential Characteristic & Output Differential Characteristic\\ 
 \hline
 1&0000, 0100, 0010, 0100& 0010, 1010, 0101, 1010\\
 \hline
 2&0000, 0100, 1001, 0100 & 0010, 1010, 0101, 1010\\
 \hline
 3&0000, 0111, 0000, 0100 & 0101, 1101, 1001, 0011\\
 \hline
 4&0000, 1011, 0010, 0100 & 0010, 1010, 0101, 1010\\
 \hline
 5&0000, 1011, 1001, 0100 & 0010, 1010, 0101, 1010\\
 \hline
 6&0000, 1110, 0000, 0100 & 0101, 1101, 1001, 0011\\
 \hline
 7&1100, 1100, 1000, 0000 & 0110, 0001, 1110, 0110\\
 \hline
\end{tabular}
\end{adjustbox}
\end{table}
According to these characteristics \cite{vahi2020separ}, we can offer the following theorem: \newline
\textbf{Theorem}: Any four consecutive b16 differential characteristic within a SEPAR Enc-block has a minimum of $7+i, 3 \leq i \leq 5$ active S-boxes.\\
\textbf{Proof}: With regarding the different values of i, we can distinguish 3 different cases: \newline
\begin{itemize}
    \item Case i = 3: with regarding the structure of selected S-boxes, middle b16s could have either 2 or 1 active S-boxes. Therefore, we have  $3+2+1+4$ or $3+1+2+4$ active S-Boxes that in either way we have 10 active S-boxes.
    \item Case i = 4: For this case, middle b16s, both have 2 active S-boxes. Another option is  1 or 3 active S-boxes respectively. Thus,there are $3+2+2+4$ or $3+1+3+4$ or $3+3+1+4$ active S-boxes which in this case, the number of active S-boxes is 11.
    \item 	Case i = 5: Reasoning for this case is similar to those for the $i=3$ and $i =4$.The computed number of active S-boxes, in this case, is 12.
\end{itemize}

Considering this theory, any differential characteristic of the SEPAR cipher, which is consists of eight Enc-blocks, must have at least 80 active S-boxes. The maximum differential probability for a  SEPAR S-box is $2^{-2}$, therefore we can conclude that in SEPAR cipher including 8 consecutive Enc-block,the differential characteristic probability is bounded by $(2^{-2})^{80}= 2^{-160}$.Therefore,  We can practically confirm that employing 8 Enc-blocks consist of 4 consecutive b16, resists the algorithm against the differential attack.

\section{Conclusion}

In this paper, we examined differential cryptanalysis of the SEPAR lightweight cryptographic algorithm. In this study, in addition to reviewing the previously offered method, compute differentials of Enc-block for a different number of b16, we analyzed the structure of Enc-block in order to compute the number of active S-boxes. This result-in the theorem that employing 4 consecutive b16 within an Enc-block, and eight Enc-blocks in the encryption algorithm, resist cipher against differential cryptanalysis and we practically confirm that the bound on the number of active S-boxes is tight.






\begin{thebibliography}{1}

\bibitem{vahi2020separ} 
Vahi, Arsalan and Jafarali Jassbi, Somaye. 
\textit{SEPAR: A New Lightweight Hybrid Encryption Algorithm with a Novel Design Approach for IoT}. 
Wireless Personal Communications 114 (2020): 2283-2314

\bibitem{jithendra2020new} 
Jithendra, K. B., and T. Kassim Shahana.
\textit{New Results in Reduced Round AES-256 Impossible Differential Cryptanalysis}. 
International Journal of Computing and Digital Systems 9.4 (2020): 755-764.

\bibitem{poschmann2009lightweight} 
Poschmann, A.
\textit{Lightweight Cryptography-Cryptographic Engineering for a Pervasive World. Ph. D. Thesis, Ruhr University Bochum, 2009.}. 
(2009)

\bibitem{bansod2017boron} 
Bansod, Gaurav, Narayan Pisharoty, and Abhijit Patil.
\textit{BORON: an ultra-lightweight and low power encryption design for pervasive computing}. 
Frontiers of Information Technology \& Electronic Engineering 18.3 (2017) : 317-331.

\bibitem{sajadieh2017new} 
Sajadieh, Mahdi, et al.
\textit{A new counting method to bound the number of active S-boxes in Rijndael and 3D}. 
Designs, Codes and Cryptography 83.2 (2017): 327-343.

\bibitem{sasaki2017new} 
Sasaki, Yu, and Yosuke Todo.
\textit{New algorithm for modeling S-box in MILP based differential and division trail search}. 
International Conference for Information Technology and Communications. Springer, Cham, 2017.


\bibitem{heys2002tutorial} 
Heys, Howard M.
\textit{A tutorial on linear and differential cryptanalysis.}. 
Cryptologia 26.3 (2002): 189-221.

\bibitem{cui2012lower} 
Cui, Ting, and Chenhui Jin.
\textit{Lower Bounds of Differential and Linear Active S-boxes for Generalized Feistel Network with SP Type F-function.}. 
Journal of Networks 7.2 (2012): 282.

\bibitem{leander2007classification} 
Leander, Gregor, and Axel Poschmann..
\textit{ On the classification of 4 bit s-boxes.}. 
International Workshop on the Arithmetic of Finite Fields. Springer, Berlin, Heidelberg, 2007.

\bibitem{saarinen2011cryptographic} 
Saarinen, Markku-Juhani O.
\textit{ Cryptographic analysis of all 4$\times$ 4-bit S-boxes}. 
International Workshop on Selected Areas in Cryptography. Springer, Berlin, Heidelberg, 2011.



\end{thebibliography}
%

\end{document}